\documentclass[%
 preprint,
 amsmath,amssymb,
 aps,
]{revtex4-1}
\usepackage[utf8]{inputenc}
\usepackage{hyperref}
\hypersetup{
    colorlinks=true,
    linkcolor=blue,
    filecolor=magenta,      
    urlcolor=blue,
    citecolor=blue,
}
\usepackage{dcolumn}
\usepackage{bm}
\usepackage{natbib}
\usepackage{float}
\usepackage[normalem]{ulem}
\usepackage{color}
\usepackage{geometry}
\usepackage[caption=false,listofformat=subsimple,labelformat=simple]{subfig}

\usepackage{graphicx}
\setcitestyle{square,numbers}
\begin{document}


\title{Geometrical Frustration and Cluster Spin Glass with  Random Graphs}

\author{Alexandre Silveira}
\author{R. Erichsen Jr.}
\author{S. G. Magalhães}
\affiliation{Instituto de Física, Universidade Federal do Rio Grande do Sul, Caixa Postal 15051,91501-970 Porto Alegre, RS, Brazil}




\date{\today}

\begin{abstract}

We 
develop a novel method based in the sparse random graph to account
the 
interplay between geometric frustration and disorder 
in cluster magnetism. 
Our theory allows to introduce the cluster network connectivity as a controllable parameter.
 Two types of  inner cluster geometry are considered: triangular and tetrahedral. The theory was developed for a general, non-uniform intra-cluster interactions, but in the present paper the results presented correspond to uniform, anti-ferromagnetic (AF) intra-clusters interactions $J_{0}/J$. 
The clusters are represented by nodes on a finite connectivity random graph, and the inter-cluster interactions are random Gaussian distributed.
The graph realizations are treated in replica theory using the formalism of order parameter functions, which allows to calculate the distribution of local fields and, as a consequence, the relevant observable. 
In the case of triangular cluster geometry, there is the onset of a classical Spin Liquid state at a temperature $T^{*}/J$ and then, a Cluster Spin Glass (CSG) phase at a temperature $T_{f}/J$. The CSG ground state is robust even for very weak disorder or large negative $J_{0}/J$. These results does not depend on the network connectivity. Nevertheless, variations in the connectivity strongly affect the level of frustration $f_{p}=-\Theta_{CW}/T_{f}$ for large $J_{0}/J$. In contrast, for the non-frustrated tetrahedral cluster geometry, the CSG ground state is suppressed for weak disorder or large negative $J_{0}/J$. The CSG boundary phase presents  a re-entrance which is dependent on the network connectivity.

%

\end{abstract}

\maketitle


\section{\label{sec:intro}Introduction}

Magnetism in clusters of spins is a very promising novel frontier both in terms of fundamental physics and applications \cite{Furrer2013}.
The starting point is that the spins cluster is a structure with constituent elements such as inner geometry, chemical composition and size, whose combined effects 
 makes cluster magnetism full of possibilities for new magnetic materials.

One of them, still little explored, is 
the relationship between cluster magnetism 
and geometrical frustration (GF) \cite{ramirez}. 
This concept has been a central topic 
in condensed matter physics \cite{book}. 
More recently, its presence has been understood 
as the cornerstone for the existence of exotic states of matter, such 
as classical or quantum spin-liquids (SL) \cite{Balents_SL}. 
It should be noted that this relationship may 
have also technological interest, since 
frustration may be behind a significant increase 
in the magnetocaloric effect \cite{Pakhira,Zhitomirsky}. 

The  question is: how to contact 
cluster magnetism with GF? 
Our  answer is based on 
the assumption that, in cluster magnetism, the spin lattice is replaced by a spin cluster network.
From that point, if there is disorder also present, we follow the main assumptions of the cluster mean field (CMF) theory
for spin glasses 
\cite{soukoulis1,soukoulis2}.  
Firstly,  one consider 
that  spin clusters themselves can interact rather than individual spins.
This is a situation similar to what has already been proposed for nanomagnetism (see, for instance, \cite{Bedanta_2008}).
Next, 
the problem is separated into intra and inter-cluster parts which are self consistently coupled leading to the glassy instability. The effects of the GF can come from the intra-cluster part considering, for instance,  a suitable inner cluster geometry with, for instance, anti-ferromagnetic (AF) interactions.

Indeed, such approach allowed to obtain non-trivial results \cite{Schmidt2015416,Zimmer2014A,JPCM_Schmidt}.
For instance, in the case of Ising spins using the kagome geometry,
it is introduced a novel mechanism to stabilize a cluster spin glass (CSG) phase at low temperature
with much weaker disorder  compared to that one required for individual spins. This mechanism is related to
the formation of a region with a classical spin-liquid (SL) 
given by a plateau in the entropy at lower temperatures
which precedes the CSG instability. We highlight that this behaviour is observed in real systems (see \cite{zheng2012,PhysRevB.79.224407,SSC_Sampa}). 
 
Nevertheless, the CMF theory assumes that the cluster network is fully connected. Such assumption may lead to an inadequate account for the GF within the self-consistent procedure which couples the intra and inter-cluster parts. 
Thus, one should evaluate properly whether and how the
variation in the connectivity of the cluster network can affect the self-consistency and, therefore, the GF effects in the disordered cluster magnetism. 

In this paper, we propose to study of the CSG state in a random graph architecture with finite connectivity. The main reason to adopt this architecture is that, since it allows to control the network connectivity, it is more realistic than the fully connected network mean field approach. We apply it for triangular and tetrahedral clusters with uniform AF intra-cluster interactions, in order to compare results with and without GF effects.

In order to deal with the realizations of the random network, we use the replica theory \cite{monassonksat, rubemwkt}, that consist in rewriting the replicated partition function in terms of order parameter functions and then taking the replica limit. The order parameter functions are then parameterized in terms of a local field distribution, which is self consistently calculated through a population dynamics algorithm. It is well known that bellow the spin glass transition, replica symmetry is broken. The instability of the replica symmetry solution is obtained through the two-replica method \cite{tworeplica}.

There are other methods to deal with the finite connectivity random network, such as the cavity method \cite{parisibethe}. The equations of the cavity method are written down considering an unique realization of the network disorder, but it becomes equivalent to the replica method when the disorder average is performed. Here, we focus mainly in the replica symmetry (RS) theory. The development of a replica symmetry breaking (RSB) theory for the cluster network with finite connectivity, to deal with geometric frustration effects, is beyond the objective of this paper.

The paper is organized as follows. In Sec. \ref{sec:method} we derive the equations for the model using the replica method. In Sec. \ref{sec:resdisc} we present the results characterizing the system by drawing phase diagrams for the thermodynamic phases. The conclusions and other remarks are found in section \ref{sec:ccr}.

\section{\label{sec:method} Model and Replica Procedure}

We study a system of $N_{cl}$ interacting clusters. The Ising spins that belong to each cluster can assume the values $\sigma_{\mu i}=\pm \frac{1}{2}$, where $\mu=1\dots N_{cl}$ is the cluster index and $i=1\dots p$ is the intra-cluster index.  The inter-cluster interaction $J_{\mu\nu}$ is chosen from a Gaussian distribution with mean zero and variance $J^{2}$. The Hamiltonian of this system is given by 
\begin{align} 
H(\boldsymbol{\sigma})= - \frac{1}{\sqrt{c}}\sum_{\mu<\nu}c_{\mu\nu}J_{\mu\nu} \sigma_{\mu}\sigma_{\nu} -\theta\sum_\mu\sigma_\mu+ \sum_{\mu}H_{0}(\boldsymbol{\sigma}_{\mu})\, 
\label{hamiltonian1} 
\end{align}
where $\boldsymbol{\sigma}\equiv\{\sigma_{\mu i}\}$ represents the state of the entire system, $\sigma_{\mu}=\sum_{i=1}^{p}\sigma_{\mu i}$ and $\boldsymbol{\sigma}_\mu$ are, respectively, total spin and state-vector of cluster $\mu$ and $\theta$ is an uniform external field coupled to the total spin of each cluster. The term $H_{0}(\boldsymbol{\sigma}_{\mu})=-\frac{1}{2}\sum_{i\neq   j}J_{ij}\sigma_{\mu i}\sigma_{\mu j}$ accounts for the intra-cluster couplings. It should be noted that, at this point, no particular choice is made for $J_{ij}$.

The elements of the connectivity matrix between clusters, $c_{\mu\nu}$, are chosen from a binary probability distribution 
\begin{align} 
p\left(c_{\mu\nu}\right) = \frac{c}{N_{cl}}\delta_{c_{\mu\nu},1} + \Big(1-\frac{c}{N_{cl}}\Big)\delta_{c_{\mu\nu},0}\, 
\label{cdistr} 
\end{align}
where the constant $c$ represents the average connectivity.

The replica method will be used to average over the quenched disorder. The disorder-averaged free energy can be written as
\begin{align}
    f(\beta)=-\lim_{\substack{N_{cl}\rightarrow\infty \\ n\rightarrow 0}} \frac{1}{\beta nN_{cl}} \log\langle {Z^n}\rangle\,,
\label{free}
\end{align}
where
\begin{equation}
    Z^n=\sum_{\boldsymbol{\sigma}^1\cdots\boldsymbol{\sigma}^n} \mathrm{e}^{-\beta \sum_\alpha H(\boldsymbol{\sigma}^\alpha)}\,
    \label{part1}
\end{equation}
is the replicated partition function, with $\alpha=1\dots n$ being the replica index. 
Averaging over the connectivity disorder this becomes, in the limit $c/N_{cl}\rightarrow 0$,
\begin{align}
    \langle Z^{n} \rangle = \sum_{\boldsymbol{\sigma}^1\cdots\boldsymbol{\sigma}^n} \exp\Big[-\beta\sum_{\alpha\mu} H_{0}(\boldsymbol{\sigma}_\mu^\alpha) + \beta\theta\sum_{\alpha\mu}\sigma_\mu^\alpha + \frac{c}{2N_{cl}} \sum_{\mu\neq \nu}\Big\langle \mathrm{e}^{\frac{\beta}{\sqrt{c}}J_{\mu\nu}\sum_{\alpha} \sigma_\mu^\alpha\sigma_\nu^\alpha}-1\Big\rangle_{J_{\mu\nu}}\Big]\,.
\label{fpart}
\end{align}

Now we start to reduce to the problem of one cluster. In the following we punctuate only the main steps and refer to Appendix (\ref{app1}) for details. The order function in problems with finite connectivity is the probability to find the replica state vector in a given state $\mathbf{s}$ \cite{monassonksat,Monasson},
\begin{align}
    P(\mathbf{s}) =\frac{1}{N_{cl}}\sum_\mu\delta_{\mathbf{s}\boldsymbol{\sigma}_\mu}\,.  \label{orderfunc}
\end{align}
Introducing Eq. (\ref{orderfunc}) in Eq. (\ref{fpart}) the partition function becomes 
\begin{align}
\nonumber 
    \big\langle Z^n \big\rangle =\int \prod_{\mathbf{s}} & dP(\mathbf{s}) d\hat{P}(\mathbf{s}) \exp N_{cl} \Big\{\sum_{\mathbf{s}} \hat{P}(\mathbf{s})P(\mathbf{s}) + \log\sum_{\mathbf{s}}\exp\Big[ -\beta\sum_\alpha H_0(\mathbf{s}^\alpha) \\ & + \beta\theta\sum_\alpha s_\alpha -\hat{P}(\mathbf{s})\Big] 
     + \frac{c}{2} \sum_{\mathbf{s}\mathbf{s}'} P(\mathbf{s})P(\mathbf{s}') \Big\langle\mathrm{e}^{\frac{\beta J} {\sqrt{c}} \sum_\alpha s_\alpha s'_\alpha} - 1\Big\rangle\Big\}\,,
  \label{fpart3}
\end{align}
where $\hat{P}(\mathbf{s})$ is an auxiliary variable and $s_\alpha=\sum_{i=1}^p s_{\alpha i}$. In the limit $N_{cl}\rightarrow\infty$, the integral in this equation can be solved through the saddle-point method. Eliminating $\hat{P}(\mathbf{s})$ through the saddle-point equations, the averaged per-cluster free energy becomes
\begin{align}
  \label{freen3}
    f(\beta) & =-\lim_{n\rightarrow 0} \frac{1}{\beta n}\mathrm{Extr} \Big\{-\frac{c}{2} \sum_{\mathbf{s}\mathbf{s}'} P(\mathbf{s}) P(\mathbf{s}') \Big\langle\mathrm{e}^{\frac{\beta J}{\sqrt{c}}\sum_\alpha s_\alpha s'_\alpha} - 1\Big\rangle \\ & + \log\sum_{\mathbf{s}} \exp\Big[ - \beta\sum_\alpha H_0(\mathbf{s}^\alpha) + \beta\theta\sum_\alpha s_\alpha + c\sum_{\mathbf{s}'} P(\mathbf{s}') \Big\langle\mathrm{e}^{\frac{\beta J}{\sqrt{c}} \sum_\alpha s_\alpha s'_\alpha} - 1 \Big\rangle\Big]\Big\}\,.  \nonumber
\end{align}
where $\mathrm{Extr}$ means to take the extreme of the expression between braces relatively to $P(\mathbf{s})$. We look for solutions satisfying the replica symmetry {\sl Ansatz} (RS), where $P(\mathbf{s})$ remains unchanged under permutation of the replica index. Since we assume that the clusters interact through their total spin, the RS {\sl Ansatz} can be written in the form 
\begin{align}
  P(\mathbf{s})=\int d\mathbf{h}\, W(\mathbf{h}) \dfrac{\exp\Big[-\beta\sum_\alpha H_0(\mathbf{s}^\alpha) + \beta\mathbf{h} \cdot\sum_\alpha \mathbf{M}(s_\alpha)\Big]}
  {\Big\{\sum_{\mathbf{s}}\exp\Big[-\beta H_0(\mathbf{s}) +
      \beta\mathbf{h}\cdot\mathbf{M}(s)\Big]\Big\}^n}\,.
\label{RSansatz}
\end{align}
Here, $\mathbf{h}$ and $\mathbf{M}(s)\equiv(s{,}s^2{,}\dots{,}s^p)$
are vectors with $p$ components, where the superscript in each components amounts to a exponent. 
A $p$-spin cluster has $p+1$ total spin states $s$. The component $i$ of vector $\mathbf{h}$ is coupled to $s^i$, allowing $\mathbf{h}$ to control the population of the $p+1$ states, while the term $H_0$ takes account of the intra-cluster states.

All the properties of the system are accessible upon knowledge of the vector-field distribution $W(\mathbf{h})$. The RS solution reads 
\begin{align}
  \label{RS8}
    W(\mathbf{h}) = \sum_kP_k \int \prod_{l=1}^k d\mathbf{h}_l\, W(\mathbf{h}_l) \Big\langle\prod_{i=1}^p \delta\Big(h_i - \sum_l\phi_i(\mathbf{h}_l{,}J_l)\Big) \Big\rangle_{J_l}\,,
\end{align}
where $h_i$ in the r.h.s. are components of vector $\mathbf{h}$ in the l.h.s. and $\phi_i(\mathbf{h}_l{,}J_l)$, $i=1,\dots p$, are functions dependent on the size of the cluster. For details about the development of Eq. (\ref{RS8}) and calculation of $\phi_i(\mathbf{h}_l{,}J_l)$ for $p=3$ and $p=4$, see the appendices.
This equation can be solved recursively through a population dynamics algorithm, to be described below.

After to obtain $W(\mathbf{h})$ it is possible to calculate the observable. For example, the per cluster magnetization, the spin-glass order parameter and the occupation number are given, respectively, by
\begin{align}
    m=\int d\mathbf{h}\, W(\mathbf{h})\langle s\rangle\,,
  \label{magnet}
\end{align}
\begin{align}
    q=\int d\mathbf{h}\, W(\mathbf{h})\langle s\rangle^2
  \label{qea}
\end{align}
and
\begin{align}
    Q=\int d\mathbf{h}\, W(\mathbf{h})\langle s^2\rangle\,,
  \label{sg}
\end{align}
where
\begin{align}
    \langle s\rangle=\dfrac{\sum_{\mathbf{s}} s\exp\big[-\beta H_0(\mathbf{s}) + \beta\mathbf{h}\cdot\mathbf{M}(s)\big]}{\sum_{\mathbf{s}} \exp\big[-\beta H_0(\mathbf{s}) + \beta\mathbf{h}\cdot\mathbf{M}(s)\big]} \,,
  \label{sigmaav}
\end{align}
and
\begin{align}
    \langle s^2\rangle=\dfrac{\sum_{\mathbf{s}} s^2\exp\big[-\beta H_0(\mathbf{s}) + \beta\mathbf{h}\cdot\mathbf{M}(s)\big]}{\sum_{\mathbf{s}} \exp\big[-\beta H_0(\mathbf{s}) + \beta\mathbf{h}\cdot\mathbf{M}(s)\big]} \,.
  \label{sigma2av}
\end{align}

To obtain the free-energy density in the RS approach, we introduce the RS {\it Ansatz} in the free-energy density, Eq. (\ref{freen3}). In the limit $n\rightarrow 0$ this results
\begin{align}
    f(\beta) & = \frac{c}{2\beta}\int d\mathbf{h}\,d\mathbf{h}' W(\mathbf{h})W(\mathbf{h}') \\ &
    \nonumber\times\dfrac{\Big\langle\sum_{\mathbf{s}\mathbf{s}'}\exp\Big[-\beta H_0(\mathbf{s}) + \beta\mathbf{h}\cdot\mathbf{M}(s) - \beta H_0(\mathbf{s}') + \beta\mathbf{h}'\cdot\mathbf{M}(s') +\beta \frac{J}{\sqrt{c}}ss'\Big]\Big\rangle_J} {\sum_{\mathbf{s}}\exp\Big[-\beta H_0(\mathbf{s}) + \beta\mathbf{h}\cdot\mathbf{M}(s)\Big]\sum_{\mathbf{s}'}\exp\Big[- \beta H_0(\mathbf{s}') + \beta\mathbf{h}'\cdot\mathbf{M}(s')\Big]} \\ 
    & \quad\quad\nonumber - \frac{1}{\beta}\sum_k P_k\int\prod_l d\mathbf{h}_l W(\mathbf{h}_l) \log\sum_{\mathbf{s}} \exp\Big[-\beta H_0(\mathbf{s}) + \beta\theta s \Big] \\ 
    & \quad\quad\quad\quad \times\nonumber\prod_l \frac{\Big\langle\sum_{\mathbf{s}_l} \exp\Big[-\beta H_0(\mathbf{s}_l) + \beta\mathbf{h}_l\cdot\mathbf{M}(s_l) + \beta \frac{J_l}{\sqrt{c}} ss_l\Big]\Big\rangle_{J_l}}{\sum_{\mathbf{s}}\exp\Big[-\beta H_0(\mathbf{s}) + \beta\mathbf{h}_l\cdot\mathbf{M}(s)\Big]}\,.  \nonumber
\end{align}

\subsection*{\label{sec:msc} Linear Magnetic Susceptibility}

The linear magnetic susceptibility $\chi=(\partial m/\partial\theta)_{\theta\rightarrow 0}=-(\partial^2f/\partial\theta^2)_{\theta\rightarrow 0}$ plays a central role in characterizing geometrical frustration \cite{ramirez}. From Eq. (\ref{freen3}) we have
\begin{align}
    \frac{\partial^2f}{\partial\theta^2}=-\frac{\beta}{n} \sum_{\mathbf{s}}P(\mathbf{s}) \Big(\sum_\alpha s_\alpha\Big)^2 + \frac{\beta}{n}\Big(\sum_{\mathbf{s}} P(\mathbf{s}) \sum_\alpha s_\alpha\Big)^2\,.
\end{align}
In the limit $n\rightarrow 0$ the second term vanishes and the first can be written in terms of $q$ and $Q$, and then   
\begin{equation}
    \chi= \beta(Q-q)\,.
\label{xifinal}
\end{equation}

\subsection*{\label{sec:sors}Stability of RS solution}

The stability of the RS solution was determined by using the two replica method \cite{tworeplica}. It consists in solving the saddle-point equations for two independent systems, only coupled through the disorder realization. The Hamiltonian for the double system reads
\begin{align} 
\nonumber
H(\boldsymbol{\sigma},\boldsymbol{\tau}) & = - \frac{1}{2\sqrt{c}}\sum_{\mu\neq\nu}c_{\mu\nu}J_{\mu\nu} \sigma_{\mu}\sigma_{\nu} - \frac{1}{2\sqrt{c}}\sum_{\mu\neq\nu}c_{\mu\nu}J_{\mu\nu} \tau_{\mu}\tau_{\nu} \\ & - \theta\sum_\mu\sigma_\mu - \theta\sum_\mu\tau_\mu + \sum_{\mu}H_{0}(\boldsymbol{\sigma}_{\mu}) + \sum_{\mu}H_{0}(\boldsymbol{\tau}_{\mu})\, 
\label{hamiltonian2} 
\end{align}

The self-consistent equation for the two-replica vector field distribution is
\begin{align}
  \label{RStwo}
    W(\mathbf{h}{,}\mathbf{h'}) & = \sum_kP_k \int \prod_{l=1}^k d\mathbf{h}_l\, d\mathbf{h'}_l\, W(\mathbf{h}_l) \, W(\mathbf{h'}_l) \\ & \times\Big\langle \prod_{i=1}^p \Big[\delta\Big(h_i - \sum_l\phi_i(\mathbf{h}_l{,}J_l)\Big) \delta\Big(h'_i - \sum_l\phi_i(\mathbf{h'}_l{,}J_l)\Big)\Big] \Big\rangle_{J_l}\,,
    \nonumber
\end{align}
 which is diagonal, i.e., 
\begin{align}
    W(\mathbf{h}{,}\mathbf{h'})=W(\mathbf{h})\delta(\mathbf{h}-\mathbf{h'})
    \label{diagonal}
\end{align}
if RS solution is stable and non-diagonal otherwise. The AT line is found by calculating the overlap between two replicas,
\begin{align}
q'=\int d\mathbf{h}\,d\mathbf{h'}\,W(\mathbf{h}{,}\mathbf{h'})\langle s\rangle(\mathbf{h})\langle s\rangle(\mathbf{h'})\,.
\end{align}
From Eqs. (\ref{qea}) and (\ref{diagonal}), $q'=q$ if RS is stable, and the appearing of a bifurcation signals the AT line.

\section{\label{sec:resdisc}Results and discussions}

\begin{figure}
\centering
\subfloat{%
\label{subfig:tri}\includegraphics[width=6cm,clip]{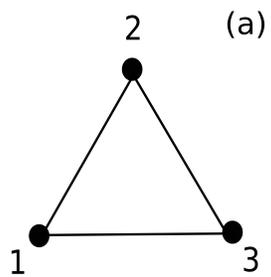}%
}\\
\subfloat{%
\label{subfig:tetra}\includegraphics[width=6cm,clip]{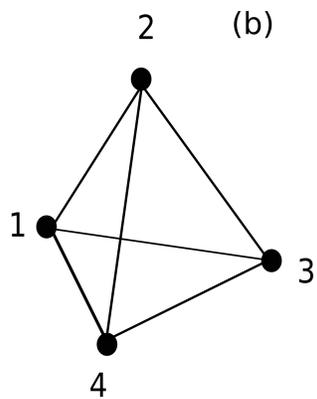}%
}

\caption{\label{fig:geome} (a) Triangular and (b) tetrahedral clusters.}
\end{figure}

Two types of cluster geometry are considered: the equilateral triangle, $p=3$ and the regular tetrahedron, $p=4$, as shown in Figs. \ref{subfig:tri} and \ref{subfig:tetra}, respectively. For simplicity, the inner couplings $J_{0}/J$ are uniform, although the theoretical framework allows to consider non-uniform couplings as well. Our analysis focuses in the interplay between the two geometries, the intra-cluster couplings $J_{0}/J$ and how it is affected by the connectivity $c$. Due to frustration effects, the most interesting is the region $J_{0}/J<0$. The standard deviation of the Gaussian distributed inter-cluster couplings $J$ is adopted as the energy scale.

The relevant observable are obtained upon the solution of the self consistent saddle point equation (\ref{RS8}) through a population dynamics algorithm \cite{parisibethe,Abou_Chacra_1973}, as follows. Initially, a population of size $\mathcal{N}$ of $p$-dimensional vector-fields is created with a certain starting guess. In each iteration, a number $k$ is chosen from a Poisson distribution of mean $c$; $k$ vector fields $\mathbf{h}_l$ and couplings $J_l$, $l=1\dots k$, are randomly chosen; the $l$-summation in each Dirac's $\delta$-function in Eq. (\ref{RS8}) is calculated. Finally, another field is randomly chosen from the population and to each of its components is assigned the corresponding $l$-summation. This procedure is repeated till the population of vector-fields converges. 

\begin{figure}
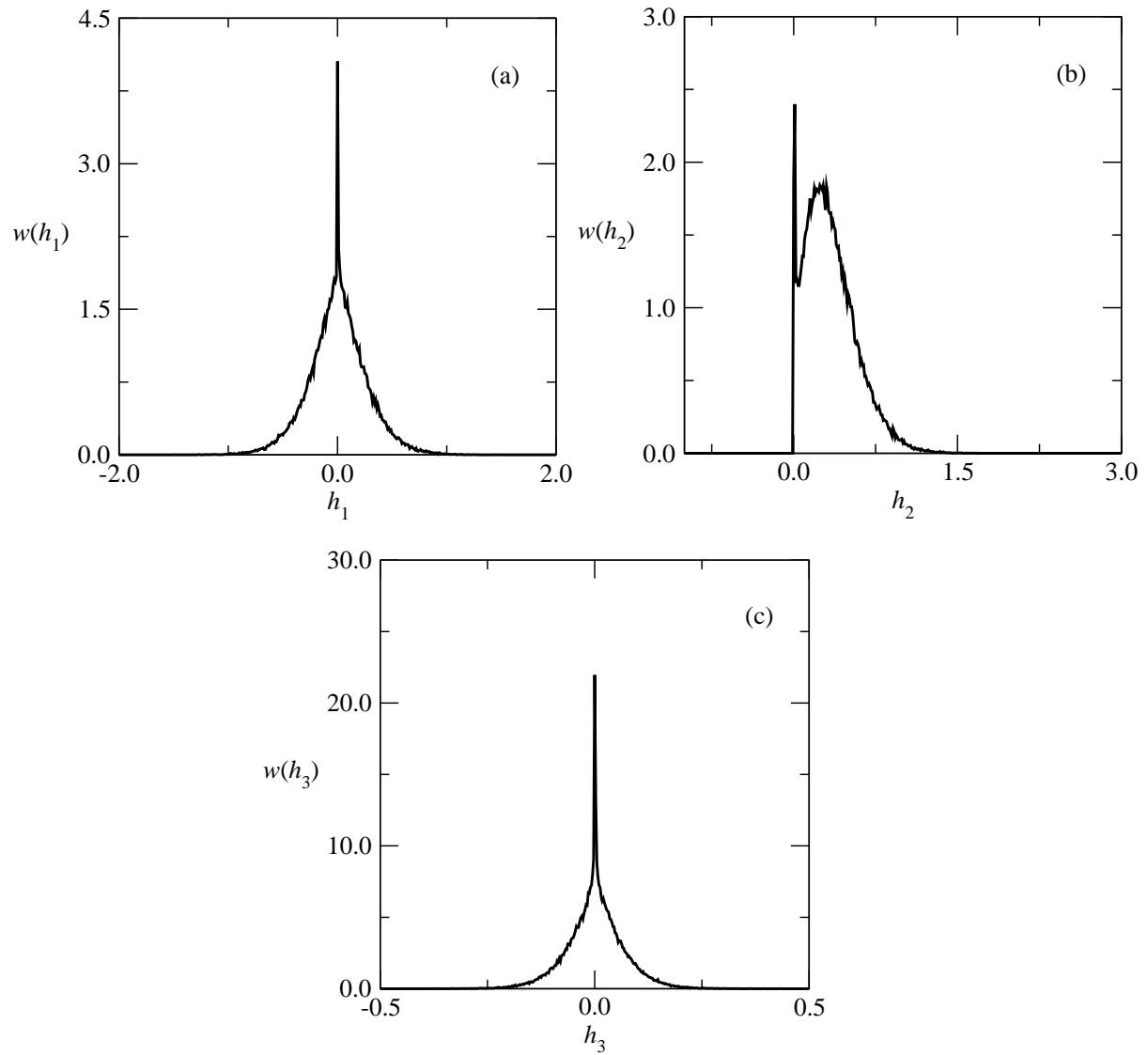

\centering
\subfloat{%
\label{subfig:wx}\includegraphics[width=8cm,clip]{wx.eps}%
}
\subfloat{%
\label{subfig:wy}\includegraphics[width=8cm,clip]{wy.eps}%
}\\
\subfloat{%
\label{subfig:wz}\includegraphics[width=8cm,clip]{wz.eps}%
}
\caption{\label{fig:wxyzj0-1} Marginal distributions for each component of a triangular cluster, for $T/J=0.1$, $c=4$ and $J_{0}/J=-5.0$, where there are strong frustration effects (see discussion below). $\mathcal{N}=10^{5}$ is the size of the population of fields. (a) Marginal along the $h_{1}$ field, (b) marginal along the $h_{2}$ field and (c) marginal along the $h_{3}$ field.}
\end{figure}

To visualize a vector-field distribution, it is convenient to use marginal distributions,
\begin{align}
    w(h_i)=\int \prod_{j\neq i} dh_j\,W(\mathbf{h})\,,
\end{align}
%
where $j$ runs over all fields but $i$.
Examples of the marginal distributions for triangular clusters are drawn in Fig. \ref{fig:wxyzj0-1}, where the system is in a CSG phase with $q > 0$ and $Q > 0$. Whenever $q > 0$, $q'\neq q$, so the RS solution is unstable into the CSG phase and, as a consequence, the AT line coincides with the CSG phase boundary. 

Next, to provide a better discussion, we present the results for each cluster geometry in two separated subsections. 

\subsection{Triangular clusters}

\begin{figure}
\centering
\includegraphics[width=8cm,clip]{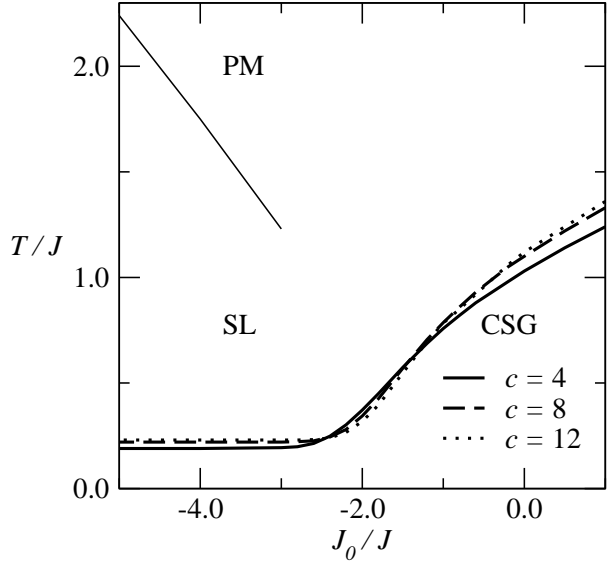}
\caption{$T/J$ versus $J_{0}/J$ phase diagrams in clusters of equilateral triangles. The thin line represents the SL to PM crossover temperature $T^{*}/J$.\label{fig:phasediagp3}}
\end{figure}

The macroscopic state is determined by calculating the order parameters $q$ and $Q$ through Eqs. (\ref{qea}) and (\ref{sg}), respectively, for a set of chosen values of $c$ and a proper range of $J_{0}/J$. To show how this unveils, Fig. \ref{fig:phasediagp3} shows the phase diagram  $T/J$ versus $J_{0}/J$ for several values of $c$. There, two types of magnetic states are observed. At high temperature, a paramagnetic (PM) phase with $Q>0$ and $q=0$ is found. Then, decreasing the temperature, there is a continuous transition to a CSG phase, with $Q>0$ and $q>0$, at the freezing temperature $T_{f}/J$. Moreover, as it will be discussed below, in a region with strong AF couplings, and therefore strong GF, there appears a crossover from the  PM phase to a classical SL state at the temperature $T^{*}/J$, at a temperature above the onset of the CSG. In particular, for $J_{0}/J<-3.0$, $T_{f}/J$ becomes independent of $J_{0}/J$, while $T^{*}/J$ becomes linearly dependent on it. Concerning the role of $c$, the phase diagram can be divided in three regions, depending on $J_{0}/J$. In the first region, with $J_{0}/J\gtrsim-1.5$, $T_{f}/J$ increases as $c$ increases. There is a second region, for $-2.5\lesssim J_{0}/J\lesssim-1.5$, $T_{f}/J$ decreases as $c$ increases. Finally, for $J_{0}/J\lesssim -2.5$, $T_{f}/J$ returns to increase with increasing $c$. It is worth to mention that in the range that was investigated, $T_{f}/J$ is weakly dependent on $c$, and it converges to the infinite connectivity behavior beyond $c\gtrsim 12$. 

\begin{figure}
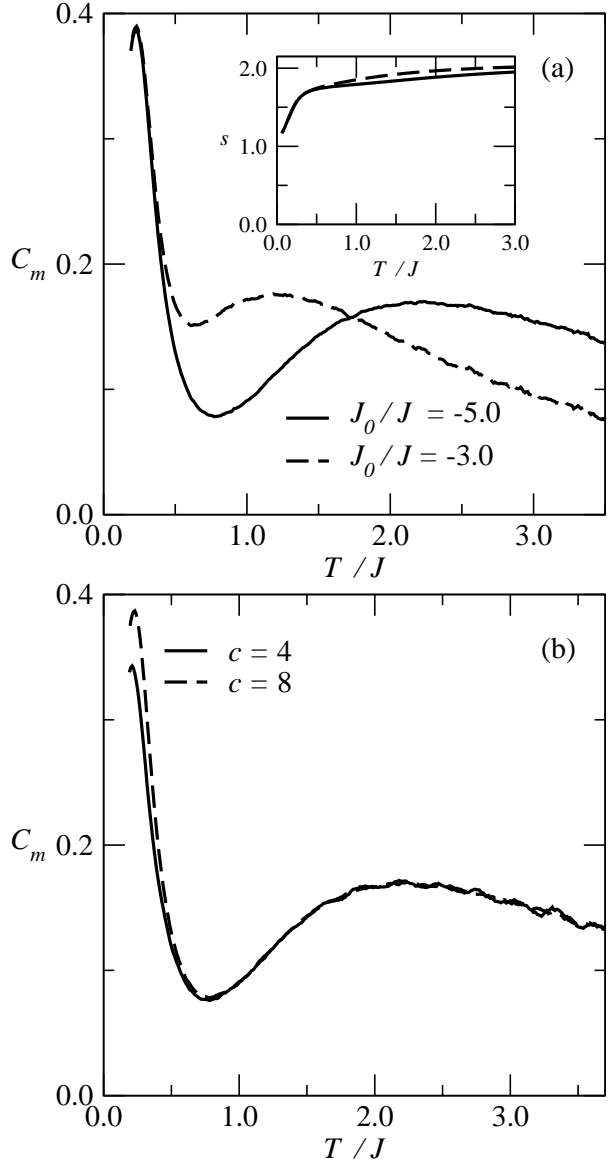

\centering
\subfloat{%
\label{subfig:sxTa}\includegraphics[width=8cm,clip]{sxT.eps}%
}
\quad\quad
\subfloat{%
\label{subfig:sxTb}\includegraphics[width=8cm,clip]{cxTc4c8.eps}%
}

\caption{(a) Specific heat $C_{m}$ vs. $T/J$ for intra-cluster couplings $J_{0}/J=-5.0$ and $J_{0}/J=-3.0$, for $c=4$. The inset shows the corresponding entropy. (b) Specific heat $C_{m}$ versus $T/J$ for $c=4$ and $c=8$, for $J_{0}/J=-3.0$. }
\end{figure}

To proceed our analysis, the entropy per cluster $s=-\partial f/\partial T$ and the magnetic specific heat $C_{m}=-T\partial^{2} f/\partial T^{2}$ were calculated. In Fig. \ref{subfig:sxTa} $s$ and $C_{m}$ are drawn for two representative values of $J_{0}/J$. The entropy plateau prior the CSG phase transition leads to a $C_{m}$ showing a two-maxima structure. There is a low temperature maximum that is relative to the loss in degrees of freedom close to the CSG transition. As the temperature continues to increase, there is a second, less pronounced maximum in $C_m$, at $T^{*}/J$. Its position varies linearly with $J_{0}/J$, as shown by the thin line in the phase diagram in Fig. \ref{fig:phasediagp3}. Strictly speaking, there is no thermodynamic transition at $T^{*}/J$, since no order parameter are going to zero there. In fact, the interval $T_{f}/J<T/J<T^{*}/J$ has always $q=0$. Nevertheless, the high temperature maximum suggests that in between the two maxima, a different paramagnetic order settles in, where internal to the clusters degrees of freedom dominate. This region corresponds to the classical SL displayed in Fig \ref{fig:phasediagp3}. Thus,  we amount $T^{*}/J$ as temperature crossover between the PM phase and the classical SL \cite{Pohle}.

To investigate the connectivity dependence on $T^{*}/J$, plots of $C_{m}$ vs. $T/J$ for two representative values of $c$ are shown in Fig. \ref{subfig:sxTb}. For $c=8$, the first $C_{m}$ maximum gets higher and is horizontally dislocated. This is consistent with a similar effect on $T_{f}/J$. Moreover, a further decrease on $c$ makes this maximum eventually disappear at the percolation limit $c=1$. The second maximum is marginally modified by changing $c$. This means that $T^{*}/J$ does not depend on the random network connectivity.

\begin{figure}
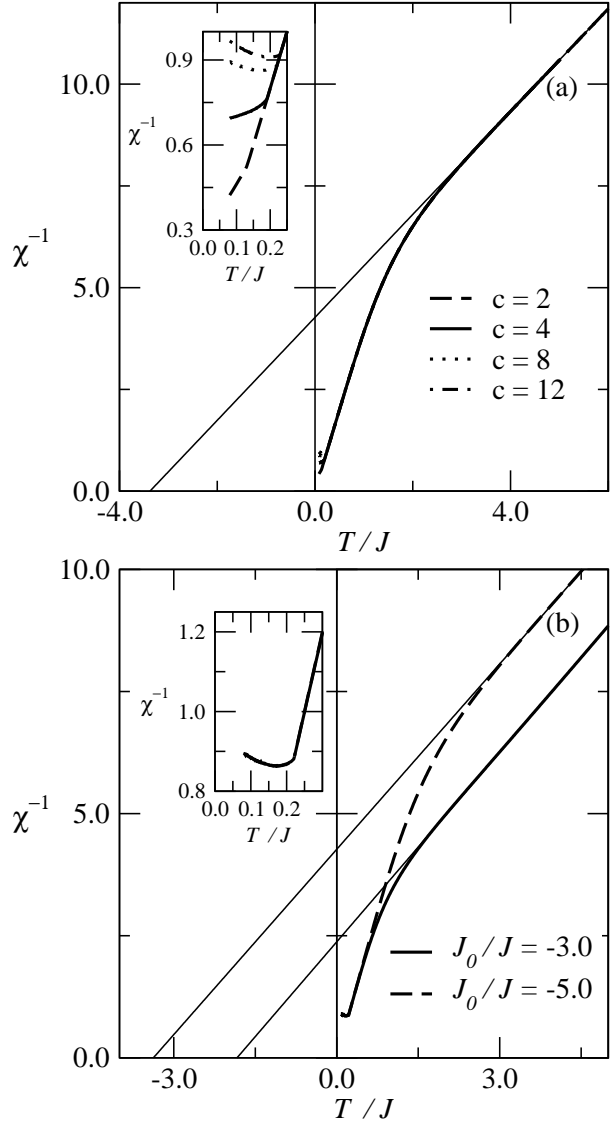

\centering
\subfloat{%
\label{subfig:chixTa}\includegraphics[width=8cm,clip]{invchic2c4c8c12.eps}%
}
\quad\quad
\subfloat{%
\label{subfig:chixTb}\includegraphics[width=8cm,clip]{cusp.eps}%
}

\caption{(a) Inverse susceptibility $\chi^{-1}$ versus $T/J$ profiles for fixed intra-cluster couplings $J_{0}/J=-5.0$ and connectivity values $c=2$, $c=4$, $c=8$ and $c=12$. The inset shows the cusps in detail. (b) Inverse susceptibility $\chi^{-1}$ vs. $T/J$ profiles for fixed $c=8$ and two values of the intra-cluster couplings: $J_{0}/J=-3.0$ and $J_{0}/J=-5.0$. The inset shows the cusps in detail. The thin straight lines represent the fit of the Curie-Weiss law to the linear region of $\chi^{-1}$.}
\end{figure}

The interplay between of $c$ and $J_{0}/J$ is also investigated regarding the level of GF  given by the parameter $f_{p}=|\Theta_{CW}|/T_{f}$ \cite{ramirez}, where $\Theta_{CW}$ is the Curie-Weiss temperature. We show in Fig. \ref{subfig:chixTa}, the inverse of susceptibility $\chi^{-1}$ displaying a cusp at $T_{f}/J$, which is characteristic of a spin glass-like transition. $\chi^{-1}$ is plotted for $c$ ranging from 2 to 12. For $T/J\gg T_{f}/J$, the behaviour of $\chi^{-1}$ does not change with an increase in $c$. However, as can be noted in Fig. \ref{subfig:chixTb},  $\Theta_{CW}/J$ is highly influenced by changing $J_{0}/J$. $\Theta_{CW}/J$ was estimated through the Curie-Weiss law $\chi(T)=C/(T-\Theta_{CW})$, from the linear region of the $\chi^{-1}$ vs. $T/J$ curves. For $J_{0}/J=-3.0$, the obtained value is $\Theta_{CW}/J\approx-1.84$ for all $c$. For $c=4$, we obtain $f_{p}\approx 9.46$, thus indicating a moderate frustrated scenario. As can be observed, $\Theta_{CW}/J$ strongly depends on $J_{0}/J$, but $f_{p}$ is also a function of $T_{f}/J$ which, by its turn, is influenced by $c$.



\begin{figure}
\centering
\includegraphics[width=8cm,clip]{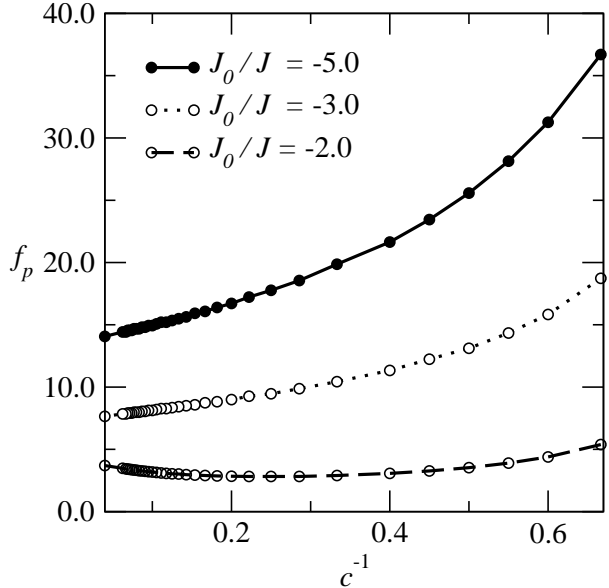}
\caption{Frustration parameter $f_{p}$ vs. $c^{-1}$.
\label{fig:frust}}
\end{figure}

To visualize how the interplay between $c$ and $J_{0}/J$ reflects on the level of frustration, curves of $f_{p}$ vs $c^{-1}$ for three values of $J_{0}/J$, are presented in Fig. \ref{fig:frust}. This allows to identify important differences in the behavior of $f_{p}$, as follows. For $J_{0}/J=-2.0$, $f_{p}$ is relatively weakly affected by the variation of $c$, but it is important to remark that there is a minimum of frustration at $c\approx 3$. This interesting point deserves further investigation to be explained. In strong contrast, for $J_{0}/J=-5.0$, $f_{p}$ presents a fast growing as $c$ decreases. This behaviour (as for $J_{0}/J=-3.0$), is coherent with the development of a SL region obtained in the phase diagram of Fig. \ref{fig:phasediagp3}. We could not go further beyond $c^{-1}\approx 0.7$ in Fig. \ref{fig:frust} because the freezing temperature $T_{f}/J$ approaches zero as $c\rightarrow 1$, causing numerical instability. Anyway, since $\Theta_{CW}/J$ is nearly $c$-independent, the $f_{p}$ parameter should diverge in this limit.

\subsection{\label{sec:cl3case}Tetrahedral clusters}

\begin{figure}
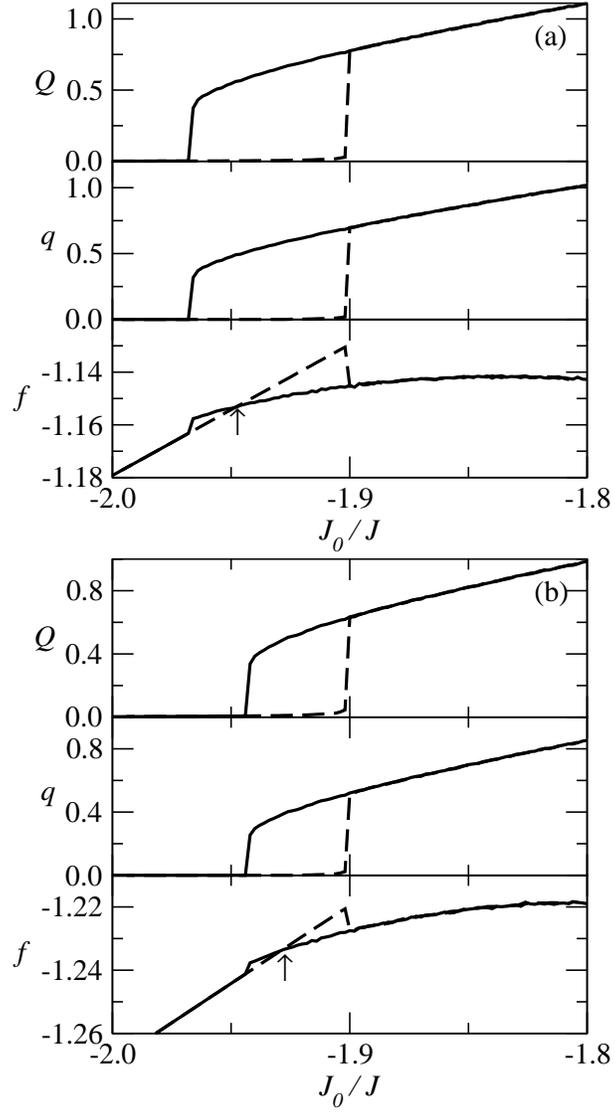

\centering
\subfloat{%
\label{subfig:fxTa}\includegraphics[width=8cm,clip]{qxJT010.eps}%
}
\quad\quad
\subfloat{%
\label{subfig:fxTb}\includegraphics[width=8cm,clip]{qxJT015.eps}%
}
\caption{(a) Total spin of the cluster $Q$, CSG order parameter $q$ and free energy $f$ vs. $J_{0}/J$ for $T/J=0.10$ and $c=8$. (b) The same, but for $T/J=0.15$. Solid (dashed) lines indicate heating (cooling). The arrows indicate the loci of the discontinuous transitions.\label{f_QxT}}
\end{figure}

There are two major differences relative to the triangular clusters.
Firstly, the cluster is not prone to geometrical frustration.
Secondly, the cluster can be fully compensated, i.e., there exist states where the total spin of the cluster $Q$ is zero. 
For instance, for $J_{0}/J\ll 0$ the 
full compensation is favoured,
as will be discussed below. Another consequence is that, contrary to triangular clusters, 
discontinuous transitions do appear for sufficiently large $c$. We refer to Figure \ref{f_QxT} as an example of the curves for 
$Q$ and 
$q$ versus $J_{0}/J$ in the vicinity of a first order CSG-PM transition, for $c=8$. The free energy $f$, that allows to localise the first order  transition, is also shown. 

\begin{figure}
\centering
\includegraphics[width=9cm,clip]{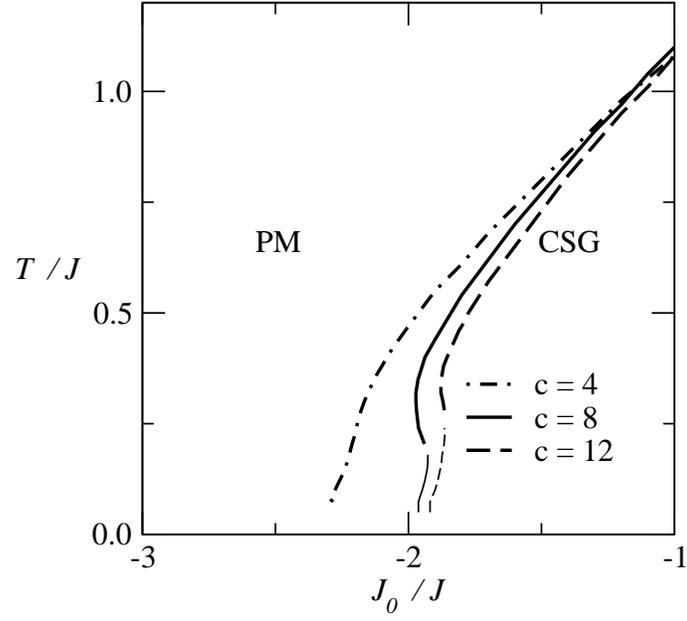}
\caption{$T/J$ vs. $J_{0}/J$ phase diagrams in clusters of regular tetrahedrons for $c=4$, $c=8$ and $c=12$. Thin lines represent first order phase transitions.\label{TvsJ0}}
\end{figure}

The general behaviour of tetrahedral clusters can be resumed in the phase diagram of Fig. \ref{TvsJ0}. The most interesting regime is $J_{0}/J<0$. For $J_{0}/J>0$ the clusters become frozen at its maximum $Q$ value.
As stated above, for $J_{0}/J\ll 0$, the fundamental state is fully compensate. This means that the total cluster spin assumes the state $s=0$, favouring a non-magnetic state, with $q=0=Q$. 
As $J_{0}/J$ become less negative, the ground state becomes the CSG phase. This occurs at $J_{0}/J\geq -1.95$ and is due to the increase in relevance of the long range random interactions. 

As shown in Figure \ref{TvsJ0}, there is a remarkable influence of the connectivity between clusters on the phase diagrams. For low connectivity, as $c=4$, the PM $\rightarrow$ CSG phase transition is always continuous. Contrary, already for $c=8$, a more complex picture appears, with a low-temperature discontinuous transition, a high-temperature continuous transition and a tricritical point between them. For $c=8$ and $c=12$, the tricritical point are located at $T_{c}/J=0.18$ and $T_{c}/J=0.24$, respectively.
%

Associated to the discontinuous PM $\rightarrow$ CSG transitions, Fig. \ref{TvsJ0} shows a reentrant region. At the transition, the CSG phase co-exists with the 
PM phase with small $Q$ 
\cite{Silva2012}. In the reentrant region the system passes from PM to CSG and then to PM. This re-entrance is similar to the unusual inverse freezing phase transition observed in non-Ising classical  models \cite{Schupper2005,Morais2012}, and is related to the full compensation of the total cluster spin at large negative $J_{0}/J$. At low temperature, the fully compensated, non-magnetic cluster state is favoured. This becomes no longer true when thermal fluctuations increase, and the long-range, disordered interaction, acts to stabilize a CSG phase. Since this is a collective effect, it depends on the connectivity. Although the re-entrance,
the principle of monotonic increase of entropy with temperature is not violated, since the small $Q$ phase has lower entropy than the CSG one that is at higher temperature. 


\section{\label{sec:ccr}Concluding remarks}

In this paper, we developed a novel theoretical framework to take into account geometric frustration  effects within the cluster magnetism with disorder in a sparse random network, where the connectivity between clusters is a controllable parameter. We have used a triangle and a tetrahedral for choices of the inner cluster structure. The intra-cluster interaction for both cases is anti-ferromagnetic (AF). 

Our results show that, for the triangular cluster geometry, the Cluster Spin Glass ground state appears even for very small disorder. Most importantly, such result does not depend on the cluster network connectivity. Furthermore,  geometric frustration effects lead to a classical spin-liquid region which is dependent on the strength of intra-cluster AF interaction $J_{0}/J$ but is very weakly dependent of the connectivity. On the other hand,
variations in the connectivity strongly affects the level of frustration $f_{p}=-\Theta_{CW}/T_f$ for larger value of $J_{0}/J$. This behaviour is particularly intense for larger values of $J_{0}/J$. Thus, the increase of $f_{p}$ is consistent with the development of the SL region. 

In the case of the tetragonal cluster geometry there appears, at low temperature, the unusual phase transition known as inverse freezing \cite{Schupper2005}, that is similar to those results already obtained for the fully connect network of clusters (see, for instance, Ref. \cite{Silva2012}). Our results show that, in the case of a low connectivity, the inverse freezing disappears, which indicates that this unusual phase transition is connectivity dependent.

To summarize, the relationship between cluster magnetism, disorder and geometric frustration was studied. For this purpose, we introduced a novel methodology in which the cluster network connectivity is a controllable parameter of the theory. We investigate two particular types of cluster geometry. However, we  do believe that the method can be extended to other kinds of cluster geometry, as for instance, the kagome one. We are currently investigating this point. Furthermore, although in this paper the intra-cluster couplings were uniform, the theory can be applied to the non-uniform case.

\section*{\label{sec:ack}Acknowledgments}

The authors would like to thank Prof. Mateus Schmidt for discussions. The present study was supported by the Brazilian agencies Conselho Nacional de Desenvolvimento Cient\'{\i}fico e Tecnol\'ogico (CNPq), Coordena\c c\~ao de Aperfei\c coamento de Pessoal de N\'{\i}vel Superior (CAPES) and Funda\c c\~ao de Amparo \`a pesquisa do Estado do RS (FAPERGS).

\appendix
\section{derivation of the saddle-point equations\label{app1}}

In order to reduce to the problem of one cluster, the first step is to withdraw the total spin of the clusters from the inner exponential in Eq. \ref{fpart}, using the identity
\begin{align}
    1=\sum_{\mathbf{s}}\prod_{\alpha=1}^n\prod_{i=1}^p \delta_{s_i^\alpha\sigma_{\mu i}^\alpha}=\sum_{\mathbf{s}}
    \delta_{\mathbf{s}\boldsymbol{\sigma}_\mu}\,,
\end{align}
where $\delta_{s\sigma}$ is the Kronecker's delta and $\mathbf{s}\equiv\{s_i^\alpha\}$ is an auxiliary spin vector. 

The replicated partition function becomes
\begin{align}
\label{fpartn}
    \langle Z^{n} \rangle = \sum_{\boldsymbol{\sigma}^1\cdots\boldsymbol{\sigma}^n} \exp\Big[& -\beta\sum_{\alpha\mu} H_{0}(\boldsymbol{\sigma}_\mu^\alpha) + \beta\theta\sum_{\alpha\mu}\sigma_\mu^\alpha \\ 
    & + \frac{c}{2N_{cl}} \sum_{\mu\neq \nu}\sum_{\mathbf{s} \mathbf{s}'}\delta_{\mathbf{s} \boldsymbol{\sigma}_{\mu}}\delta_{\mathbf{s}' \boldsymbol{\sigma}_{\nu}}\Big\langle\mathrm{e}^{\frac{\beta J}{\sqrt{c}} \sum_{\alpha}s_\alpha s'_\alpha}-1\Big\rangle_J\Big]\,,
    \nonumber
\end{align}
where $s_\alpha=\sum_{i=1}^p s_i^\alpha$. 

The order function, Eq. (\ref{orderfunc}),
is introduced in Eq. (\ref{fpart}) through a Dirac's delta function and then, using the integral representation for the Dirac's delta function, the replicated partition function becomes
\begin{align}
    \nonumber \big\langle Z^n\big\rangle & =\sum_{\boldsymbol{\sigma}^1\cdots\boldsymbol{\sigma}^n} \int\prod_{\mathbf{s}}dP(\mathbf{s}) d\hat{P}(\mathbf{s})\exp\Big\{- \beta\sum_{\alpha\mu} H_0(\boldsymbol{\sigma}_\mu^\alpha) + \beta\theta\sum_{\alpha\mu}\sigma_\mu^\alpha + \sum_{\mathbf{s}} \hat{P}(\mathbf{s})P(\mathbf{s}) \\ & -\frac{1}{N_{cl}}\sum_{\mathbf{s}} \hat{P}(\mathbf{s})\sum_\mu\delta_{\mathbf{s}\boldsymbol{\sigma}_\mu} + \frac{cN_{cl}}{2} \sum_{\mathbf{s}\mathbf{s}'} P(\mathbf{s})P(\mathbf{s}') \Big\langle\mathrm{e}^{\frac{\beta J}{\sqrt{c}}\sum_\alpha s_\alpha s'_\alpha}-1\Big\rangle_J \Big\}\,.  
\label{fpart2}
\end{align}

The summation over the spin variables $\mbox{\boldmath$\sigma$}_\mu^\alpha
=(\sigma_{\mu 1}^\alpha\dots\sigma_{\mu p}^\alpha)$ concerns the factor
\begin{align}
    \%=\sum_{\boldsymbol{\sigma}^1\cdots\boldsymbol{\sigma}^n} \exp\Big[-\beta\sum_{\alpha\mu} H_0(\boldsymbol{\sigma}_\mu^\alpha) + \beta\theta\sum_{\alpha\mu}\sigma_\mu^\alpha -\frac{1}{N_{cl}}\sum_{\mathbf{s}} \hat{P}(\mathbf{s})\sum_\mu\delta_{\mathbf{s}\boldsymbol{\sigma}_\mu}\Big]\,. \label{trace1}
\end{align}
Summing over $\mathbf{s}$ in the second term of the exponential and rearranging terms, this becomes
\begin{align}
    \%=\prod_\mu\sum_{\boldsymbol{\sigma}_\mu} \exp\Big[-\beta\sum_\alpha H_0(\boldsymbol{\sigma}_\mu^\alpha) + \beta\theta\sum_\alpha\sigma_\mu^\alpha -\frac{1}{N_{cl}}\hat{P}(\boldsymbol{\sigma}_\mu)\Big]\,. 
  \label{trace2}
\end{align}
Since the clusters are decoupled, this reduces to the problem of one cluster, and can be expressed as
\begin{align}
    \%=\exp N_{cl}\log\sum_{\mathbf{s}} \exp\Big[-\beta\sum_\alpha H_0(\mathbf{s}^\alpha) + \beta\theta\sum_\alpha s_\alpha -\frac{1}{N_{cl}}\hat{P}(\mathbf{s})\Big]\,.
\label{trace3}
\end{align}
Changing variables $\hat{P}\rightarrow N_{cl}\hat{P}$, we obtain Eq. (\ref{fpart3}).

Introducing Eq. (\ref{fpart3}) in Eq. (\ref{free}), the averaged per-cluster free energy reads
\begin{align}
\nonumber 
    f(\beta) =-\lim_{n\rightarrow 0}\frac{1}{\beta n}\mathrm{Extr}\Big\{\sum_{\mathbf{s}} \hat{P}(\mathbf{s})P(\mathbf{s}) & + \log\sum_{\mathbf{s}}\exp\Big[- \beta\sum_\alpha H_0(\mathbf{s}^\alpha) + \beta\theta\sum_\alpha s_\alpha -\hat{P}(\mathbf{s})\Big] \\ &+ \frac{c}{2} \sum_{\mathbf{s\mathbf{s}}'} P(\mathbf{s})P(\mathbf{s}') \Big\langle\mathrm{e}^{\frac{\beta    J}{\sqrt{c}}\sum_\alpha s_\alpha s'_\alpha}-1\Big\rangle \Big\}\,, \label{freen2}
\end{align}
where $\mathrm{Extr}$ means to take the extreme of the expression between braces relatively to variables $\hat{P}(\mathbf{s})$ and $P(\mathbf{s})$. This is imposed by the saddle-point equations
\begin{align}
    \frac{\partial f(\beta)}{\partial\hat{P}(\mathbf{s})} = 0 = \frac{\partial f(\beta)}{\partial P(\mathbf{s})}\,,
\label{saddle0}
\end{align}
which are written as
\begin{align}
    P(\mathbf{s})=\dfrac{\exp\Big[ -\beta\sum_\alpha H_0(\mathbf{s}^\alpha) + \beta\theta\sum_\alpha s_\alpha - \hat{P}(\mathbf{s})\Big]} {\sum_{\mathbf{s}'}\exp\Big[-\beta\sum_\alpha H_0(\mathbf{s}'^\alpha) + \beta\theta\sum_\alpha s'_\alpha -\hat{P}(\mathbf{s}')\Big]}
  \label{saddle1}
\end{align}
and
\begin{align}
    \hat{P}(\mathbf{s})=-c\sum_{\mathbf{s}'} P(\mathbf{s}')\big\langle\mathrm{e}^{\frac{\beta J}{\sqrt{c}} \sum_\alpha s_\alpha s'_\alpha}-1\big\rangle\,.
  \label{saddle2}
\end{align}
Eliminating $\hat{P}(\mathbf{s})$ in Eq. (\ref{saddle1}), we obtain a self-consistent equation for $P(\mathbf{s})$:
\begin{align}
    P(\mathbf{s})=\dfrac{\exp\Big[- \beta\sum_\alpha H_0(\mathbf{s}^\alpha) + \beta\theta\sum_\alpha s_\alpha +c\sum_{\mathbf{s}'} P(\mathbf{s}')\Big\langle\mathrm{e}^{\frac{\beta J}{\sqrt{c}}\sum_\alpha s_\alpha s'_\alpha}-1\Big\rangle\Big]} {\sum_{\mathbf{s}'}\exp\Big[- \beta\sum_\alpha H_0(\mathbf{s}'^\alpha) + \beta\theta\sum_\alpha s'_\alpha + c\sum_{\mathbf{s}''} P(\mathbf{s}'')\Big\langle\mathrm{e}^{\frac{\beta J}{\sqrt{c}}\sum_\alpha s'_\alpha s''_\alpha} - 1\Big\rangle\Big]}\,.
  \label{saddle3}
\end{align}
Introducing Eqs. (\ref{saddle2}) and (\ref{saddle3}) in Eq. (\ref{freen2}) we obtain the per cluster free-energy, Eq. (\ref{freen3}).

To obtain the vector local-field distribution, Eq. (\ref{RS8}), we introduce the RS {\sl Ansatz}, Eq. (\ref{RSansatz}), in the saddle-point equation, Eq. (\ref{saddle3}). Since $P(\mathbf{s})$ is a probability, the denominator on the r.h.s. of Eq. (\ref{saddle3}) is equal to 1. Expanding the exponential on the numerator, we have
\begin{align}
\label{RS4}
    P(\mathbf{s}) = \exp & \Big[- \beta\sum_\alpha H_0(\mathbf{s}^\alpha) + \beta\theta\sum_\alpha s_\alpha\Big] \sum_kP_k \prod_{l=1}^k\sum_{\mathbf{s}_l} \int d\mathbf{h}_l\,W(\mathbf{h}_l) \\
    & \times\frac{\Big\langle\exp\Big[-\beta\sum_\alpha H_0(\mathbf{s}^\alpha_l) + \beta\mathbf{h}_l\cdot\sum_\alpha\mathbf{M}(s_{\alpha l}) + \beta \frac{J_l}{\sqrt{c}} \sum_\alpha s_\alpha s_{\alpha l}\Big]\Big\rangle_{J_l}}{\Big\{\sum_{\mathbf{s}} \exp\Big[-\beta H_0(\mathbf{s}) + \beta\mathbf{h}_l\cdot\mathbf{M}(s)\Big]\Big\}^n}\,,
  \nonumber
\end{align}
where $ P_k=\sum_k\mathrm{e}^{-c}c^k/k!$ is a Poissonian weight. Rearranging terms, this can be rewritten as
\begin{align}
  \label{RS41}
  P(\mathbf{s}) = \exp\Big[- \beta\sum_\alpha H_0(\mathbf{s}^\alpha) + \beta\theta\sum_\alpha s_\alpha\Big] &\sum_kP_k \int \prod_{l=1}^k d\mathbf{h}_l\,W(\mathbf{h}_l)\\
  &\times \dfrac{\Big\langle\exp\sum_\alpha \sum_ s\delta_{ss_\alpha}\log\chi_s(\mathbf{h}_l{,}J_l)\Big\rangle_{J_l}}{\chi_0^n (\mathbf{h}_l,0)}\,,
  \nonumber
  \end{align}
where
\begin{align}
  \chi_s(\mathbf{h}{,}J)=\sum_{\mathbf{s}'}\exp\Big[-\beta H_0(\mathbf{s}') + \beta\mathbf{h}\cdot\mathbf{M}(s') + \beta \frac{J}{\sqrt{c}}ss'\Big]
\end{align}
and the Kroenecker's delta was introduced to factorize the replica index $\alpha$.
If $s$ and $s_\alpha$ are $p+1$--state spin variables, $\delta_{ss_\alpha}$ is a symmetric polynomial in powers of $s$ and $s_\alpha$. 
Introducing the corresponding Kroenecker's delta and summing over the spin variables $\mathbf{s}$, Eq. (\ref{RS41}) can be written in the form
\begin{align}
\label{RS5}
  & P(\mathbf{s}) = \sum_kP_k \int \prod_{l=1}^k d\mathbf{h}_l\,W(\mathbf{h}_l) \\ 
  &\quad \times \dfrac{\Big\langle\exp\Big[- \beta\sum_\alpha H_0(\mathbf{s}^\alpha) + \beta n\sum_l\phi_0(\mathbf{h}_l{,}J_l) + \beta \sum_l\boldsymbol{\phi}(\mathbf{h}_l{,}J_l) \cdot\sum_\alpha\mathbf{M}(s_\alpha)\Big] \Big\rangle_{J_l}}{\chi_0^n (\mathbf{h}_l{,}0)}\,.
  \nonumber
\end{align}
Here, $\boldsymbol{\phi}(\mathbf{h}{,}J)$ denotes a $p$--component vector whose components are the functions $\phi_i(\mathbf{h}{,}J)$. The zeroth component need not to be calculated, since at the end we will take the limit $n\rightarrow 0$ . A detailed calculation, as well as the derivation of the vector $\boldsymbol{\phi}(\mathbf{h}{,}J)$ for triangular and tetragonal lattices will be shown in Appendices \ref{app2} and \ref{app3}, respectively. Introducing the RS ``Ansatz'' in the l.h.s. and considering that the denominator in the r.h.s. goes to 1, Eq. (\ref{RS5}) becomes
\begin{align}
  \label{RS6}
  \int d\mathbf{h}\, W(\mathbf{h}) \exp\Big[ - \beta\sum_\alpha & H_0(\mathbf{s}^\alpha) + \beta\mathbf{h}\cdot\sum_\alpha\mathbf{M}(s_\alpha)\Big] \\
   = \sum_kP_k \int \prod_{l=1}^k & d\mathbf{h}_l\,W(\mathbf{h}_l)\Big\langle \exp\Big[- \beta\sum_\alpha H_0(\mathbf{s}^\alpha) \nonumber \\ 
  & 
  \quad\quad + \beta\sum_l\boldsymbol{\phi} (\mathbf{h}_l{,}J_l) \cdot\sum_\alpha\mathbf{M}(s_\alpha)\Big]
  \Big\rangle_{J_l}\,.
  \nonumber
\end{align}
Introducing a Dirac's delta function for each component of the field vector $\mathbf{h}$ in the r.h.s. of Eq. (\ref{RS6}) we have
\begin{align}
  \label{RS7}
  \int d\mathbf{h}\, W(\mathbf{h}) \exp\Big[- & \beta \sum_\alpha H_0(\mathbf{s}^\alpha) + \beta\mathbf{h}\cdot\sum_\alpha\mathbf{M}(s_\alpha)\Big] \\
   = \int d\mathbf{h} \sum_k & P_k \int \prod_{l=1}^k d\mathbf{h}_l\,W(\mathbf{h}_l)\Big\langle\prod_{i=1}^p \delta\Big(h_i - \sum_l\phi_i(\mathbf{h}_l{,}J_l)\Big)\Big\rangle_{J_l} \nonumber \\
  & \quad\quad\quad\quad\quad\times\exp\Big[- \beta\sum_\alpha H_0(\mathbf{s}^\alpha) 
  + \beta\mathbf{h} \cdot\sum_\alpha\mathbf{M}(s_\alpha)\Big]\,.
  \nonumber
\end{align}
Comparing both sides of Eq. (\ref{RS7}) we obtain Eq. (\ref{RS8}).


\section{3-spins clusters \label{app2}}

The cluster spin assumes four states: $-3/2{,}-1/2{,}1/2{,}3/2$. The four-state Kroenecker's delta reads
\begin{align}
    \delta_{ss_\alpha}=\frac{41}{64}-\frac{5}{16}\big(s^2 + s_\alpha^2\big) + \frac{365}{144}ss_\alpha - \frac{41}{36}\big(s^3 s_\alpha + ss_\alpha^3\big) + \frac{1}{4}s^2s_\alpha^2 + \frac{5}{9}s^3 s_\alpha^3\,.
  \label{delta4}
\end{align}

Introducing Eq. (\ref{delta4}) in Eq. (\ref{RS41}), summing over $s$ and rearranging terms we obtain Eq. (\ref{RS5}), with
\begin{align}
    \beta\phi_1(\mathbf{h}{,}J) = \frac{\theta}{k} + \frac{27}{24} \log\frac{\chi_{\frac{1}{2}}(\mathbf{h}{,}J)} {\chi_{-\frac{1}{2}}(\mathbf{h}{,}J)} - \frac{1}{24} \log\frac{\chi_{\frac{3}{2}}(\mathbf{h}{,}J)}{\chi_{-\frac{3}{2}}(\mathbf{h}{,}J)}\,,
\end{align}
\begin{align}
    \beta\phi_2(\mathbf{h}{,}J) = - \frac{1}{4} \log\chi_{\frac{1}{2}}(\mathbf{h}{,}J) \chi_{-\frac{1}{2}}(\mathbf{h}{,}J) + \frac{1}{4} \log\chi_{\frac{3}{2}}(\mathbf{h}{,}J) \chi_{-\frac{3}{2}}(\mathbf{h}{,}J)\,,
\end{align}
and
\begin{align}
    \beta\phi_3(\mathbf{h}{,}J) = - \frac{1}{2} \log\frac{\chi_{\frac{1}{2}}(\mathbf{h}{,}J)} {\chi_{-\frac{1}{2}}(\mathbf{h}{,}J)} + \frac{1}{6} \log\frac{\chi_{\frac{3}{2}}(\mathbf{h}{,}J)} {\chi_{-\frac{3}{2}}(\mathbf{h}{,}J)}\,.
\end{align}

\section{4-spins clusters \label{app3}}

The cluster spin assumes five states: $-2{,}-1{,}0{,}1{,}2$. The five-state Kroenecker's delta reads
\begin{align}
    \nonumber \delta_{ss_\alpha} =1 - \frac{5}{4}\big(s^2 + s_\alpha^2\big) & + \frac{65}{72}ss_\alpha + \frac{1}{4}\big(s^4 + s_\alpha^4\big) - \frac{17}{72}\big(s^3s_\alpha + ss_\alpha^3\big)\\ & + \frac{707}{288}s^2 s_\alpha^2 -\frac{155}{288}\big(s^4 s^2_\alpha + s^2 s_\alpha^4\big) + \frac{5}{72}s^3 s_\alpha^3 + \frac{35}{288}s^4 s_\alpha^4\,.
  \label{delta5}
\end{align}

Introducing Eq. (\ref{delta5}) in Eq. (\ref{RS41}), summing over $s$ and rearranging terms we obtain Eq. (\ref{RS5}), with
\begin{align}
    \beta\phi_1(\mathbf{h}{,}J) = \frac{\theta}{k} - \frac{1}{12} \log\frac{\chi_{+2}(\mathbf{h}{,}J)} {\chi_{-2}(\mathbf{h}{,}J)} + \frac{2}{3} \log\frac{\chi_{+1}(\mathbf{h}{,}J)} {\chi_{-1}(\mathbf{h}{,}J)}\,,
\end{align}
\begin{align}
    \beta\phi_2(\mathbf{h}{,}J) = -\frac{1}{24} \log\chi_{+2}(\mathbf{h}{,}J)\chi_{-2}(\mathbf{h}{,}J) + \frac{2}{3} \log\chi_{+1}(\mathbf{h}{,}J) \chi_{-1}(\mathbf{h}{,}J) - \frac{5}{4}\log\chi_0(\mathbf{h}{,}J)\,,
\end{align}
\begin{align}
    \beta\phi_3(\mathbf{h}{,}J) = \frac{1}{12} \log\frac{\chi_{+2}(\mathbf{h}{,}J)} {\chi_{-2}(\mathbf{h}{,}J)} - \frac{1}{6} \log\frac{\chi_{+1}(\mathbf{h}{,}J)} {\chi_{-1}(\mathbf{h}{,}J)}\,,
\end{align}
\begin{align}
    \beta\phi_4(\mathbf{h}{,}J) = \frac{1}{24} \log\chi_{+2}(\mathbf{h}{,}J)\chi_{-2}(\mathbf{h}{,}J) - \frac{1}{6} \log\chi_{+1}(\mathbf{h}{,}J) \chi_{-1}(\mathbf{h}{,}J) + \frac{1}{4}\log\chi_0(\mathbf{h}{,}J)\,.
\end{align}

\end{document}